\documentclass[nofootinbib,aps,prl,twocolumn,showpacs,superscriptaddress]{revtex4-1}

\usepackage{graphicx,xcolor} 
\usepackage{amsmath}
\usepackage{amsthm, amsmath, amssymb}
\usepackage[loose,nice]{units} 
\usepackage{mathtools} 
\usepackage{pdfpages}

\makeatletter
\AtBeginDocument{\let\LS@rot\@undefined}
\makeatother

  \def\a{\alpha} \def\b{\beta}
\def\n{\hat{\mathbf{n}}} 
\def\np{\hat{\mathbf{n}}_\perp}

\def\r{\mathbf{r}}

\newcommand{\A}[1]{#1_{\!A}}
\renewcommand{\a}{\alpha}
\renewcommand{\b}{\beta}

\newcommand{\pd}{\partial}
\newcommand{\M}{\mathcal{M}}
\newcommand{\D}{\mathcal{D}}

\begin{document}
\title{Curved geometries from planar director fields - \\
Solving the two-dimensional inverse problem}
\date{\today}
\author{Itay Griniasty} \affiliation{Department of Physics of Complex Systems,
Weizmann Institute of Science, Rehovot 76100, Israel} 
\author{Hillel Aharoni} \affiliation{Department of Physics of Complex Systems, Weizmann Institute of Science, Rehovot 76100, Israel} \affiliation{University of Pennsylvania, Philadelphia, USA}
\author{Efi Efrati} \email{efi.efrati@weizmann.ac.il} \affiliation{Department of
Physics of Complex Systems, Weizmann Institute of Science, Rehovot
76100, Israel}

\begin{abstract}	
Thin nematic elastomers, composite hydrogels and plant tissues are among many systems that display uniform anisotropic deformation upon external actuation. In these materials, the spatial orientation variation of a local director field induces intricate global shape changes. Despite extensive recent efforts, to date, there is no general solution to the inverse design problem: how to design a director field that deforms exactly into a desired surface geometry upon actuation, or whether such a field exists.  
In this work, we phrase this inverse problem as a hyperbolic system of differential equations. We prove that the inverse problem is locally integrable, provide an algorithm for its integration, and derive bounds on global solutions. We classify the set of director fields that deform into a given surface, thus paving the way to finding optimized fields.
\end{abstract} 
\maketitle

\begin{figure*}[t]
	\includegraphics[width=0.9\textwidth]{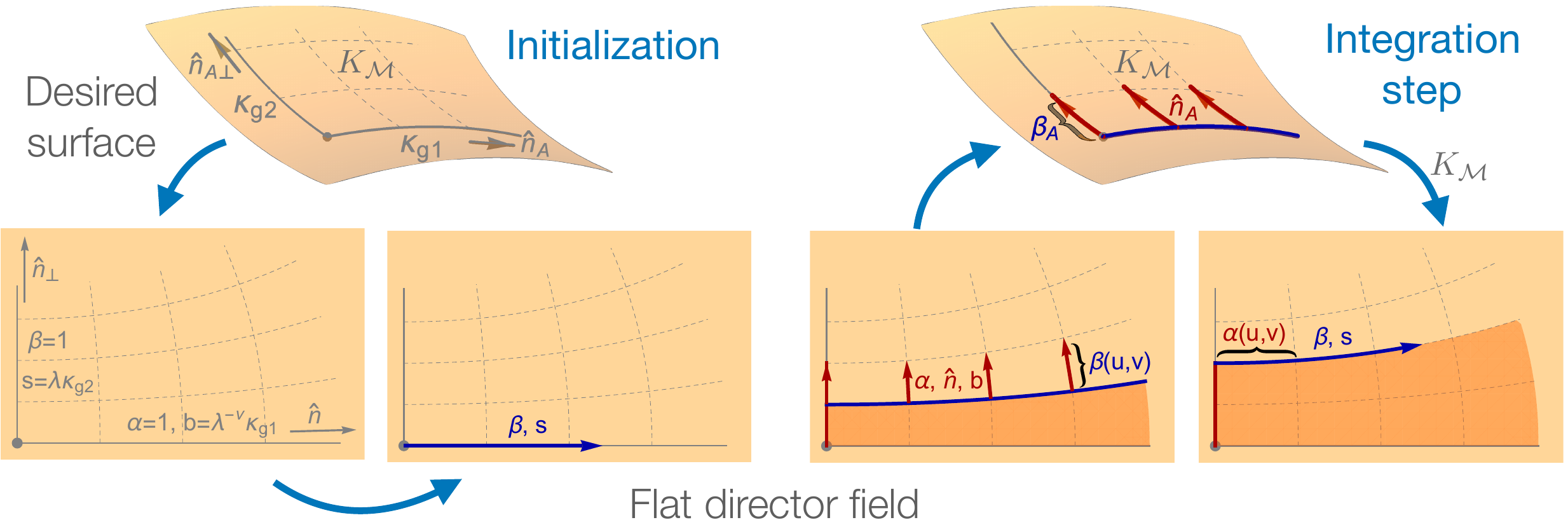}
	\caption{Integration of the director field.
			{\em Left: Initialization.} Given a surface with curvature \(K\) the initial condition consists of two perpendicular curves that will become integral curves of the actuated director and its perpendicular. The corresponding unactuated integral curves inherit the geosedic curvatures $b=\lambda^{-\nu} \kappa_{g1}$ and  $s=\lambda \kappa_{g2}$ respectively, and we are free to set $\a=1$ along the former and $\b=1$ along the latter.
				We next integrate to obtain \(\b\) and \(s\) along the director integral curve completing the initialization of the SOE.
		{\em Right: Iterative integration step.}
		(i) \(\a,\n,\r\) and \(b\) are integrated a \(dv\) step along \(\np\) on the flat sheet. \(\A{\n}\) and \(\A{\r}\) are  integrated a step $dv$ along \(\n_{A\perp}\) on the desired curved surface. (ii) \(\b\) and \(s\) are integrated along \(\n\) according to the Gaussian curvature \(K_\M\) pulled back from the desired surface. With the information at hand this step can now be reiterated. 	}
	\label{fig:eqns}
\end{figure*}

{
Many fiber-reinforced thin biological tissues \cite{FZ55,AEKS11,RM09,AAESK12} and
 synthetic sheets of responsive materials \cite{Gla16,War16,Aha18,WM17,War18} deform into their desired shapes by a uniform anisotropic deformation.  
Upon actuation these effectively 2D materials expand by a constant factor along the fibers and shrink by a different factor along the perpendicular direction. While the length variations along these principal axes are constant across the material, the spatial variation in the direction of the principal axes allows this simple mode of uniform deformation to result in rich and intricate shapes.

The fiber orientation is described by the field \(\n(\r)\) called the director. Together with the spatially constant shrinkage/expansion factors the director field uniquely defines the two-dimensional geometry that is obtained upon actuation. 
The actuation can be achieved through changing a variety of ambient conditions:
temperature or light in liquid crystal elastomers \cite{Kupfer1991,Finkelmann2001,WT03}, humidity for a variety of plants \cite{FZ55,AEKS11,RM09}, and immersion in water in fiber-reinforces hydrogels \cite{Gla16}.  

Predicting the geometry obtained upon activation as a function of the prescribed director field has been recently resolved \cite{Modes2010a,Modes2011,ASK14,Mostajeran2015}. This geometry, captured by the two dimensional Riemannian metric, 
however, does not uniquely define the obtained surface. 
A given metric will correspond to a wide and typically continuous family of surfaces. The geometric rigidity that arises from Gaussian curvature sign variations as well as imposed boundary conditions serve to narrow down this wide family. Nonetheless, selecting the desired surface among all embeddings requires some control over the principal curvatures of the surface.
Several techniques to partially control the principal curvatures of the thin sheet have been proposed and implemented \cite{ASK14,Gla16,Aha18}, yet these techniques are system-specific and depend strongly on the elastic constitutive relations. 
In what follows we only address the universal problem associated with controlling the two-dimensional Riemannian geometry. 
The desired surface is an isometric embedding of the obtained solution, however, other embeddings may exist. Selecting among these will be addressed in the future.

Recent responsive 3D printing  (often termed 4D) applications \cite{Gla16} and advances in programmable nematic elastomer production \cite{Aha18} are aimed at producing a desired surface upon actuation, and thus give rise to the inverse problem: 
{\em What is the planar director orientation field that will result in a desired geometry upon actuation?} 
There have been several recent advances in addressing the inverse problem. In \cite{ASK14,WM17, War18} director fields are found for surfaces of revolution while in \cite{Aha18} approximate solutions are found numerically for arbitrary surface geometries. In \cite{Plucinsky2016,BRRS19} the anisotropic deformation was allowed to vary spatially, leading to a less constrained inverse problem, to which possible solutions were presented. 
Alas, an exact solution to the full inverse problem was not found in the general case, nor was it shown to exist.

In this letter we formulate the inverse problem as a set of partial differential equations (PDEs). We show that the system is well posed and 
demonstrate director fields that curve into arbitrary surfaces by integrating these equations. We present an algorithm that when provided with a desired geometry and appropriate initial conditions integrates the sought director field. This approach allows us to explore the limits of director induced deformations and characterize the set of director fields that produce a desired geometry. Characterizing the collection of solutions opens the door to optimization of the choice of initial data with respect to desirable properties such as maximizing coverage or minimizing distortions. 

To find the set of equations describing the inverse problem, we first rephrase the recently solved forward problem: What is the geometry assumed when a prescribed director field is actuated. 

{{\it The forward problem. --- }
Consider an initially flat thin sheet made of a uniform material characterized by a planar director field $\n=(\cos(\theta),\sin(\theta))$. Upon actuation the material shrinks by a factor $\lambda$ along the director and expands by a factor $\lambda^{-\nu}$ along the perpendicular direction. As shown in \cite{ASK14,Mostajeran2015} the Gaussian curvature of the actuated surface reads
\[
\begin{aligned}
\A{K}=(\lambda^{2\nu}-\lambda^{-2})\times\Bigl(&
\cos(2\theta)\left((\partial_{y}\theta)^{2}-(\partial_{x}\theta)^{2}+\pd_{x}\pd_{y}\theta \right)\\
&+\frac{\sin(2\theta)}{2}\left(\pd_{x}^{2}\theta-4\pd_{x}\theta \pd_{y}\theta-\pd_{y}^{2}\theta \right) \Bigr).
\end{aligned}
\]
Solving the inverse problem amounts to finding a planar director field, $\theta(x,y)$, that satisfies the above non-linear partial differential equation. This formidable problem is further complicated as the Gaussian curvature is naturally given on the curved surface, and not in the Cartesian coordinates $(x,y)$.

Exploiting the natural coordinates and scalars that characterize the director field  \cite{NE17} we next recast the system in a form that allows explicit integration of the inverse problem.
In two-dimensions one may always define a parametrization $\r(u,v)$ such that $u$ parametric curves (along which $v$ is constant) are everywhere tangent to the director, whereas the $v$ parametric lines are perpendicular to the director, 
\begin{equation}
\pd_u\r=\a \n,\qquad
\pd_v\r=\b\np.
\label{eq:uv}
\end{equation}
The metric of the flat sheet with respect to this parametrization  is given by 
\(dl^2=\a^2du^2+\b^2dv^2\).
Upon actuation the flat sheet shrinks and expands along  \(\n\) and \(\np\) respectively.  The arc-length parameters 
thus change according to  
\begin{equation}
\A{\a}=\lambda \a ,\qquad \A{\b}=\lambda^{-\nu}\b,
\label{eq:ab}
\end{equation}
and the actuated metric remains diagonal and is given by
\(\A{dl} ^2=\A{\a}^2du^2+\A{\b}^2dv^2\).

A two dimensional director field, \(\n\), is fully characterized by two local scalar fields  \cite{NE17}; its intrinsic bend, $b$, and splay, $s$, which are given by 
\begin{equation}
 b= \np\cdot (\n\cdot\nabla)\n\,,\qquad s= \np\cdot(\n_{\perp}\cdot\nabla)\n\,.
\label{eq:bsn}
\end{equation}
Geometrically, the bend and splay represent gradients in $\n$ along $\n$ and across it, and correspond to the geodesic curvatures of  the \(u\) and \(v\) parametric curves, respectively. 
They are thus related to the flat arc-lengths by
\begin{equation}
b=-\frac{\pd_v\a}{\a \b},\qquad 
s=\frac{\pd_u\b}{\a \b}.
\label{eq:bs}
\end{equation}
Given the two dimensional metric one could express the Gaussian curvature in terms of the splay, the bend and their directional derivatives. For the case where the director is given in a planar domain this leads to 
\begin{equation}
0=s^2+\frac{1}{\a} \pd_u s+b^2 - \frac{1}{\b} \pd_v b,
\label{eq:Kflat}
\end{equation}
see \cite{NE17,SM}.
Upon actuation the metric component rescale according to \eqref{eq:ab} and the actuated Gaussian curvature reads
\begin{equation}
-\A{K}=\lambda^{-2}\left (s^2+\frac{1}{\a} \pd_u s \right )+\lambda^{2\nu}\left (b^2 - \frac{1}{\b}\pd_v b\right ).
\label{eq:K}
\end{equation}
As the actuated geometry is fully described by $\A{K}$ this completes the solution of the forward problem.  
}

{{\it The inverse problem. ---}
Given a curved surface \(\M\) with Gaussian curvature \(K_{\M}\), we seek a flat director field \(\n\) that upon actuation will assume the geometry of $\M$, and in particular $\A{K}=K_{\M}$.

Combining equations \eqref{eq:Kflat} and \eqref{eq:K} we find the propagation equations for the bend \(b\) and splay \(s\) on the flat sheet along \(\np\) and \(\n\)
\begin{align}
\frac{1}{\b}\pd_v b&=b^2-\frac{K_{\M}}{\lambda^{-2}-\lambda^{2\nu}},
\label{eq:bsuv}\\
\frac{1}{\a}\pd_u s&=-s^2-\frac{K_{\M}}{\lambda^{-2}-\lambda^{2\nu}}.\notag
\end{align}

The System Of Equations (SOE) comprised of \eqref{eq:uv},\eqref{eq:bsn},\eqref{eq:bs} and \eqref{eq:bsuv} allows us to find a parametrization of a flat sheet, \(\r(u,v)\), and a director field tangent to the \(u\)-parametric curves \(\n\propto \pd_u \r\), such that when the flat sheet is actuated it deforms into a surface with the desired curvature \(K_{\M}\). 

Not all systems of partial differential equations are solvable. To allow a solution from initial data they must satisfy integrability conditions. These estimate the predicted variation of the solution along closed paths, and must vanish. 
The integrability conditions for Eq.\eqref{eq:uv}, which propagate $\r$, are synonymous with equations \eqref{eq:bsn} and  \eqref{eq:bs}. The integrability conditions for  \eqref{eq:bsn}, which propagate $\n$, yield \eqref{eq:Kflat}. The remaining differential relations, namely \eqref{eq:bs} and \eqref{eq:bsuv}, propagate information only along one direction and thus cannot lead to contradictions in the integrated value of the solutions.   

In the particular and simple case where the desired geometry is characterized by \(K_{\M}=const\), the SOE are self-contained and can be integrated directly. This will result in a planar director field $\n$ that when actuated will adopt the geometry of constant Gaussian curvature $K_{\M}$, as can be seen in Fig.~\ref{fig:realisation} and \ref{fig:sphere}.

In contrast, when the desired geometry is characterized by a spatially varying Gaussian curvature the SOE is not self-contained, as solving equations \eqref{eq:bsuv} requires knowledge of the curvature $K_{\M}(u,v)$ at position $\A\r$. This, in turn, requires that we also know the embedding $\A\r(u,v)$. While  $\A\a,\A\b,\A b$ and $\A s$ are algebraically related to their flat counterparts, through equations \eqref{eq:ab} and \eqref{eq:bs}, this does not hold for the actuated director field $\A\n$ and the exact embedding $\A\r(u,v)$. To obtain the director $\A\n$ and embedding $\A\r(u,v)$ one has to integrate the curved versions of equations \eqref{eq:bsn} and \eqref{eq:uv} respectively.

This results in an integration scheme in which 
at every step one solves $\a,\b,b,s,\n$ and $\r(u,v)$ on the flat sheet, and then uses this information to integrate $\A\n$ and $\A\r$ to obtain $K_{\M}(u,v)$ for the next integration step, as depicted in Fig.~\ref{fig:eqns}. See Supplemental Material (SM) for more details \cite{SM}.

\begin{figure}[t]
	\centering 
	\includegraphics[width=\columnwidth]{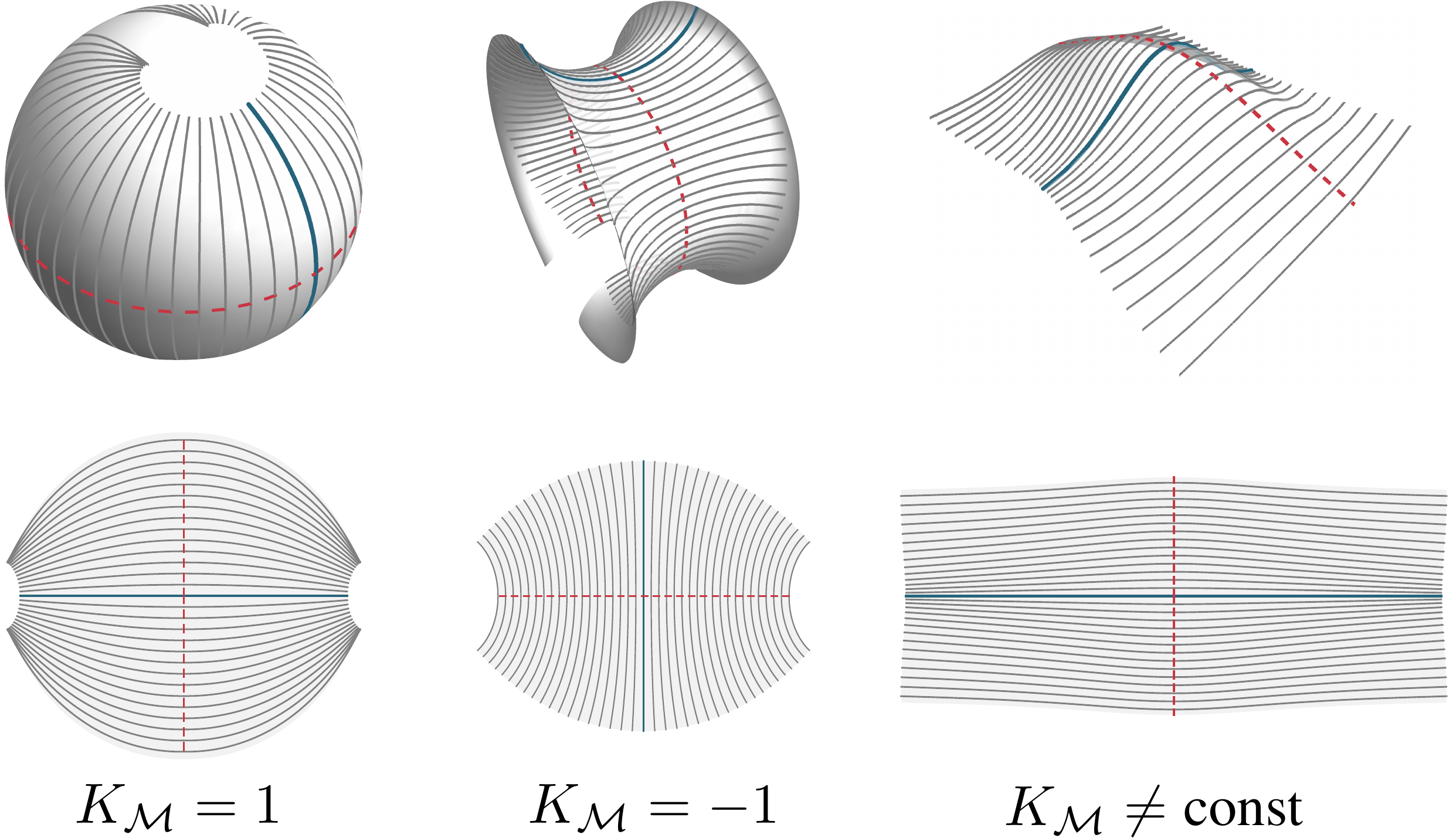}
	\caption{Solutions of the SOE: Director curves on a flat sheet and the shapes they take when actuated, found by integrating the SOE. Left: Sphere with Gaussian curvature \(K_\M=1\). Center: Constant negative-curvature, \(K_\M=-1\), body of rotation. Right: Anisotropic Gaussian surface, with varying Gaussian curvature.}
	\label{fig:realisation}
\end{figure} 

{{\it Initial conditions. --- }
Given a curved surface $\M$, solving the inverse problem amounts to finding a director field $\n$ that satisfies the SOE with respect to $\M$ in a flat domain $\mathcal{D}$. 

To solve the Cauchy problem for the SOE, i.e. to find initial conditions around which unique solutions exist, and to characterize these solutions we first need to understand the structure of the SOE. Equations \eqref{eq:Kflat} and \eqref{eq:K} form a hyperbolic set of equations for $s$ and $b$. Bringing them to their canonical form \eqref{eq:bsuv} identifies 
the characteristic lines along which information propagates with the parametric curves of $u$ and $v$. Examining equations  \eqref{eq:bs} we find a similar hyperbolic structure, and that the role of the parametric curves as carriers of partial information is preserved also for $\a$ and $\b$. Initial data for $\a$ and $b$, the arc-length and geodesic curvature of $u$-lines, is propagated along $v$-lines, while initial data for $\b$ and $s$,  the arc-length and geodesic curvature of $v$-lines, propagates along $u$-lines. Once $\a,\b,b$ and $s$ are known, $\n$ and $\r$ can be obtained by directly integrating \eqref{eq:bsn} and \eqref{eq:uv} respectively.  
The structure of the SOE is quasi-linear and is reminiscent of the equations associated with the embedding of hyperbolic surfaces in $\mathbb{R}^{3}$ \cite{Roz62,PS95}. 

This hyperbolic structure implies that prescribing $\a,\b,b$ and $s$ along a non-characteristic curve $\gamma\in\mathcal{D}$, as well as $\n$ and $\r$ at some point along the curve, leads to a unique solution in its vicinity \cite{Hadamard}. Equivalent types of initial data that could be used include prescribing  $\n$ and $\mathbf{\nabla}_{\perp}\n$  along $\gamma$, or alternatively prescribing $\n$ and $b$ along the same curve.

A particularly convenient and geometrically transparent choice for the system at hand is prescribing a Goursat initial condition  \cite{Goursat}, in which the initial data is divided into two components each given on a different line. Specifically we pick two orthogonal curves in $\M$, one of which we set to be a $u$-line (i.e. an integral curve of the director) and the other a $v$-line (an integral curve of the perpendicular to the director). We next show how to integrate the SOE from such initial conditions.
}

\begin{figure}[t]
	\centering 
	\includegraphics[width=\columnwidth]{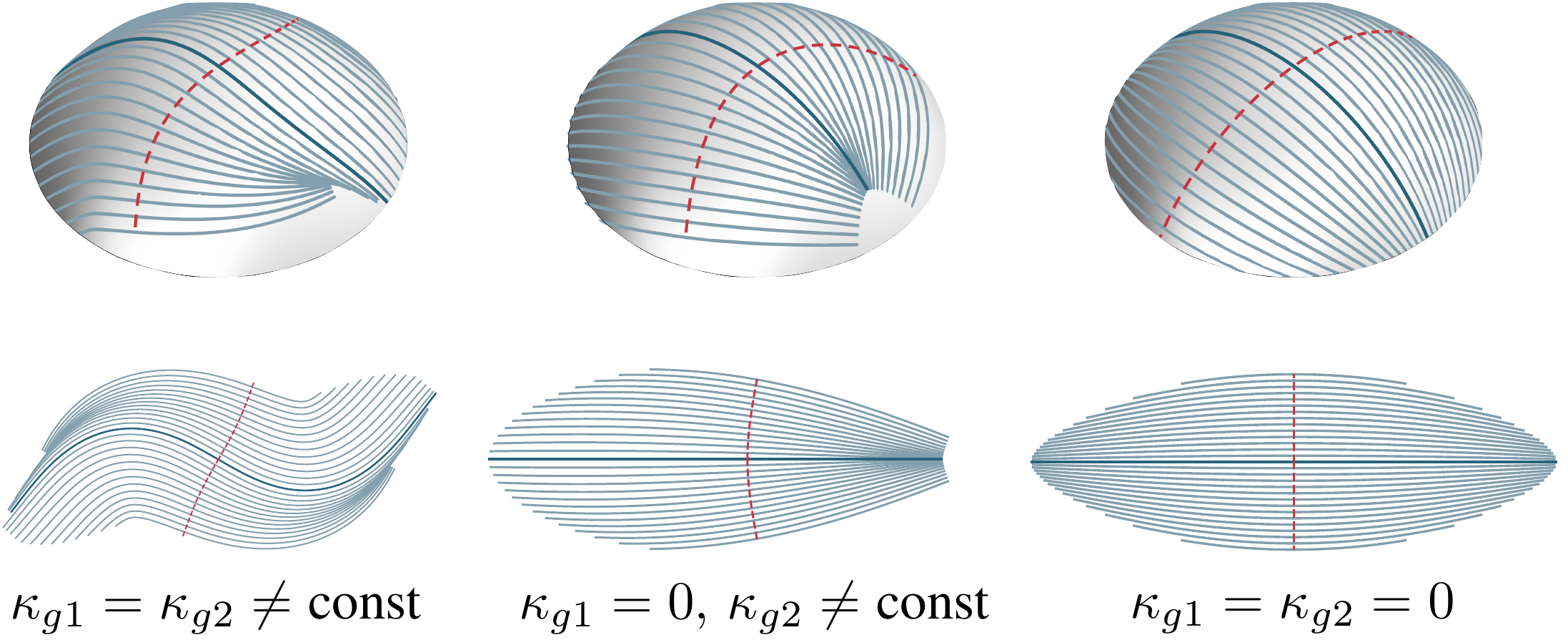}
	\caption{Distinct director fields (bottom) deforming into the same section of the unit sphere (top). The solutions are characterized by the geodesic curvatures of the initial curves, the \(u\)-base line (thick blue) with \(\kappa_{g1}\) and the \(v\)-base line (dashed red) with \(\kappa_{g2}\), from which they are integrated. }
	\label{fig:sphere}
\end{figure}

{{\it Integrating the inverse equation. ---} The solution is initialized by pulling back the two initial curves on the curved surface \(\M\), onto the flat domain \(\D\) by integrating equations \eqref{eq:uv} and \eqref{eq:bsn} (Fig.~\ref{fig:eqns} left). The bend and splay of the flat director field in  \(\D\) are algebraically related to the actuated bend and splay, which correspond to the geodesic curvature of the \(u\)-line and \(v\)-line in  \(\M\) respectively.
The arc-lengths' gauge freedom is fixed by setting \(\a\)  to \(1\) on the \(u\)-line, and \(\b\)  to \(1\) on the \(v\)-line. This completes the setting of the Goursat initial condition -- two perpendicular characteristic curves in $\D$, the $u$-baseline along which we know $\a$ and  $b$, and the $v$-baseline along which $\b$ and $s$ are known. We finish the initialization step by obtaining the values of $\b$ and $s$ along the $u$-baseline through the integration of equations \eqref{eq:bs} and ~\eqref{eq:bsuv}. 

Following the initialization, the SOE is  integrated  via a reiterated two-step tango: Knowing $\r,\n,\A\r,\A\n,\a,\b,s,b$ on a $u$-line allows us to integrate 
 \(\r,\A{\r},\n,\A\n,\a,b\) one integration step along $v$, creating the next $u$-line.  The missing information for $\b$ and $s$ does not propagate along $v$, but can now be integrated from the $v$-baseline along the newly formed $u$-line. The Gaussian curvature \(K_\M\) is inherited through the embedding \(\A{\r}(u,v)\).    This two step iteration is repeated until either the curved surface \(\M\) is covered by actuated director curves (see figure \ref{fig:realisation}), or until one reaches a singularity of the equations where \(\b=0\) or \(\a=0\), i.e. at a defect in the integrated director field.
}

\begin{figure}[t]
	\centering 
	\includegraphics[height=0.5\columnwidth]{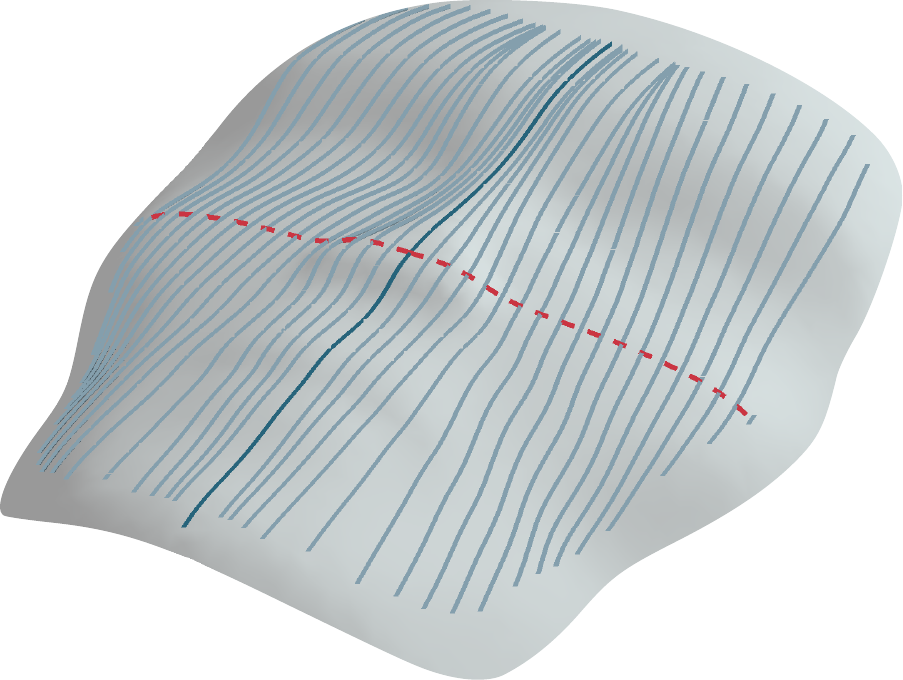}
	\includegraphics[height=0.5\columnwidth]{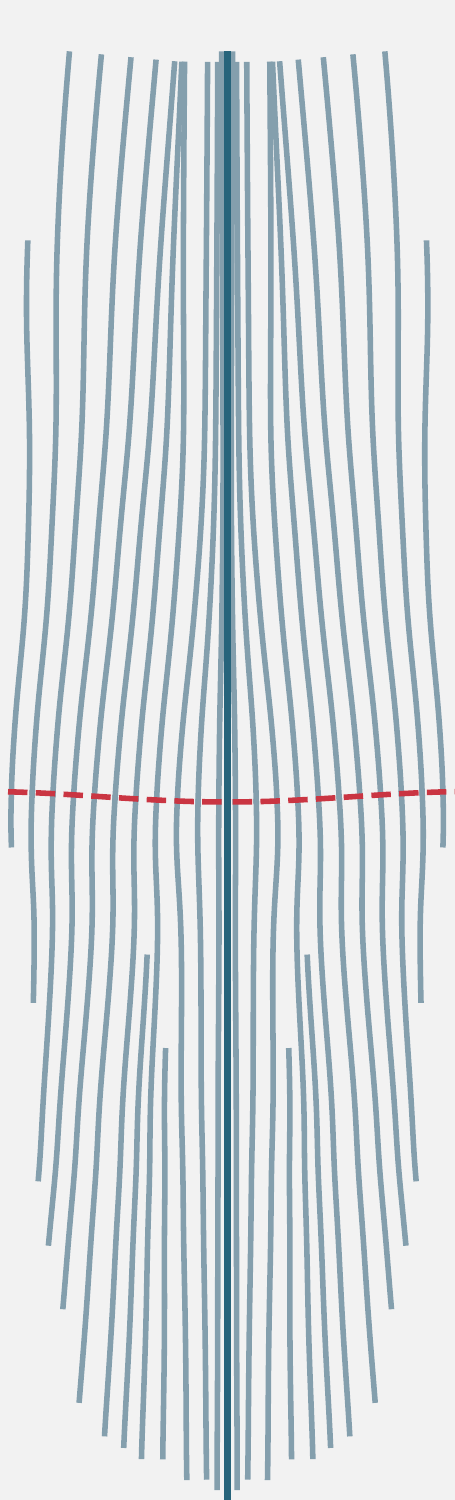}
	\caption{Surface mimicking a human face. The director field is found by integrating the SOE according to the iterative algorithm depicted in figure \ref{fig:eqns}.}
	\label{fig:face}
\end{figure}

{{\it Integration distance bounds. ---}
One naturally wonders: how far can the SOE be integrated with respect to a given curved surface before encountering a singularity? To examine this question we recast equation \eqref{eq:bsuv}  into an ordinary differential equation for \(\b\) along a $u$-line, and for \(\a\) along a $v$-line:
\begin{align}
\left.\frac{\partial^{2}\b}{\partial \sigma^{2}}\right|_{v}+\frac{K_\M}{\lambda^{-2}-\lambda^{2\nu}}\b=0, \quad\left.\frac{\partial^{2}\a}{\partial \sigma^{2}}\right|_{u}-\frac{K_\M}{\lambda^{-2}-\lambda^{2\nu}}\a=0,\notag
\end{align}
where $\sigma$ is an arc-length parameter along the respective parametric curves.
Considering a surface of positive curvature \(K_\M\ge K_0>0\), for some constant $K_{0}$, the evolution of \(\b\) on the \(u\)-characteristic is bound from above by a harmonic oscillator with a  frequency \(\omega=\sqrt{\frac{K_0}{\lambda^{-2}-\lambda^{2\nu}}}\), thus it must arrive at a singular value \(\b=0\) within a propagation distance \(x_0\le\pi/\omega\) (for a detailed analysis see SM \cite{SM}). In contrast, \(\a\) never develops a singularity in this scenario.
Such director fields thus have a finite horizon along $\n$ but can be continued indefinitely along $\np$. For example, director field can wrap around a sphere multiple times along \(\np\), while it can cover no more than half the sphere's circumference along \(\n\). For a surface of strictly negative curvature \(K_\M\le-K_0<0\), the roles of \(\n\) and \(\np\) interchange, and defects appear within a distance \(\pi/\omega\) along \(\np\). This is a manifestation of the ``orthogonal duality'' \cite{Mostajeran2015}  by which a uniform quarter-rotation to the director field, \(\n\to\np\), leads to a surface of opposite curvature, \(K\to-K\).

As shown above, certain geometries cannot be realized by actuating a defect free director field. In other cases, however, an encountered singularity may be pushed further away by varying the initial curves' geodesic curvature. Introducing carefully chosen grain boundaries to the director field could further extend the range of attainable geometries. This is somewhat analogous to  the ``lines of inflection'' introduced in \cite{Gemmer2016} to allow extending the limits of isometric embeddings of a given hyperbolic geometry. A grain boundary of a similar type in the nematic director can be seen in the numerically obtained field designed to actuate into the form of a face in \cite{Aha18}. See \cite{SM}. 
}

{{\it Discussion. ---} 	
The hyperbolic system of differential equations derived in this work allows us to establish the existence of a solution to the inverse problem, at least locally, and to explicitly show how to calculate it. For clarity we have used a flat unactuated sheet, yet the scheme holds for any unactuated geometry, see \cite{SM}.
The system of equations also predicts that near any calculated solution exist infinitely many other different solutions -- distinct director fields that correspond to the same surface geometry. These solutions are classified by the orthogonal base curves on the desired surface (see Fig.~\ref{fig:sphere} for examples). Practically this classification is given by a point and an initial direction on $\M$, as well as the geodesic curvatures of the two base curves, $b(u,0)$ and $s(0,v)$. Should a global solution to the inverse problem exist systematically exploring the possible initial conditions will allow us to find it. 

In the context of biological tissues, and in particular for fiber reinforced hygromorphing tissue \cite{FZ55,AEKS11,RM09,AAESK12}, the existence of a continuum of director fields that result in the same geometry may imply that the realized director field is optimal with respect to some biological function. Our explicit classification of these isometric textures provides a way to systematically explore this natural optimization process and possibly identify the evolutionary forces behind it. 

Similarly, for manmade systems we may exploit the freedom of initial data to optimize with respect to a desired outcome; 
Solutions with limited bend and splay might be easier to implement in the lab, while other solutions might better suit a specific target shape on account of their initial buckling, anisotropic elastic moduli, or the extrinsic curvature fields that may be imprinted onto them \cite{Aha18}. 
Moreover, one may prescribe not only the final configuration but 
guide the path the system will evolve through on its way to the final state. 
Implemented to the growing variety of responsive materials actuated via a director field -- nematic elastomer \cite{Aha18},  4D-printed fiber reinforced hydro-gels \cite{Gla16} and baro-morphing elastomeric sheets \cite{BRRS19} -- the inverse design problem solved in this work could pave the way to soft machines of unprecedented
accuracy, control and  capabilities.
}

\textbf{Acknowledgments }
We thank Ido Levin, Cyrus Mostajeran 
Eran Sharon, Shankar Venkataramani and Mark Warner for stimulating discussions. E.E. thanks the Alon fellowship and the Ernst and Kaethe Ascher foundation. I.G. is grateful to the Azrieli Foundation for the award of an Azrieli Fellowship.
This work was supported by ISF grant 1479/16 and Minerva grant 712273.
H.A. was supported by NSF grant DMR-1262047.

\bibliographystyle{apsrev4-1}	
\bibliography{gaebib.bib}

\begin{thebibliography}{25}%
\makeatletter
\providecommand \@ifxundefined [1]{%
 \@ifx{#1\undefined}
}%
\providecommand \@ifnum [1]{%
 \ifnum #1\expandafter \@firstoftwo
 \else \expandafter \@secondoftwo
 \fi
}%
\providecommand \@ifx [1]{%
 \ifx #1\expandafter \@firstoftwo
 \else \expandafter \@secondoftwo
 \fi
}%
\providecommand \natexlab [1]{#1}%
\providecommand \enquote  [1]{``#1''}%
\providecommand \bibnamefont  [1]{#1}%
\providecommand \bibfnamefont [1]{#1}%
\providecommand \citenamefont [1]{#1}%
\providecommand \href@noop [0]{\@secondoftwo}%
\providecommand \href [0]{\begingroup \@sanitize@url \@href}%
\providecommand \@href[1]{\@@startlink{#1}\@@href}%
\providecommand \@@href[1]{\endgroup#1\@@endlink}%
\providecommand \@sanitize@url [0]{\catcode `\\12\catcode `\$12\catcode
  `\&12\catcode `\#12\catcode `\^12\catcode `\_12\catcode `\%12\relax}%
\providecommand \@@startlink[1]{}%
\providecommand \@@endlink[0]{}%
\providecommand \url  [0]{\begingroup\@sanitize@url \@url }%
\providecommand \@url [1]{\endgroup\@href {#1}{\urlprefix }}%
\providecommand \urlprefix  [0]{URL }%
\providecommand \Eprint [0]{\href }%
\providecommand \doibase [0]{http://dx.doi.org/}%
\providecommand \selectlanguage [0]{\@gobble}%
\providecommand \bibinfo  [0]{\@secondoftwo}%
\providecommand \bibfield  [0]{\@secondoftwo}%
\providecommand \translation [1]{[#1]}%
\providecommand \BibitemOpen [0]{}%
\providecommand \bibitemStop [0]{}%
\providecommand \bibitemNoStop [0]{.\EOS\space}%
\providecommand \EOS [0]{\spacefactor3000\relax}%
\providecommand \BibitemShut  [1]{\csname bibitem#1\endcsname}%
\let\auto@bib@innerbib\@empty
\bibitem [{\citenamefont {Fahn}\ and\ \citenamefont {Zohary}(1955)}]{FZ55}%
  \BibitemOpen
  \bibfield  {author} {\bibinfo {author} {\bibfnamefont {A.}~\bibnamefont
  {Fahn}}\ and\ \bibinfo {author} {\bibfnamefont {M.}~\bibnamefont {Zohary}},\
  }\href@noop {} {\bibfield  {journal} {\bibinfo  {journal} {Phytomorph.}\
  }\textbf {\bibinfo {volume} {5}},\ \bibinfo {pages} {99} (\bibinfo {year}
  {1955})}\BibitemShut {NoStop}%
\bibitem [{\citenamefont {Armon}\ \emph {et~al.}(2011)\citenamefont {Armon},
  \citenamefont {Efrati}, \citenamefont {Kupferman},\ and\ \citenamefont
  {Sharon}}]{AEKS11}%
  \BibitemOpen
  \bibfield  {author} {\bibinfo {author} {\bibfnamefont {S.}~\bibnamefont
  {Armon}}, \bibinfo {author} {\bibfnamefont {E.}~\bibnamefont {Efrati}},
  \bibinfo {author} {\bibfnamefont {R.}~\bibnamefont {Kupferman}}, \ and\
  \bibinfo {author} {\bibfnamefont {E.}~\bibnamefont {Sharon}},\ }\href
  {\doibase 10.1126/science.1203874} {\bibfield  {journal} {\bibinfo  {journal}
  {Science}\ }\textbf {\bibinfo {volume} {333}},\ \bibinfo {pages} {1726}
  (\bibinfo {year} {2011})}\BibitemShut {NoStop}%
\bibitem [{\citenamefont {Reyssat}\ and\ \citenamefont
  {Mahadevan}(2009)}]{RM09}%
  \BibitemOpen
  \bibfield  {author} {\bibinfo {author} {\bibfnamefont {E.}~\bibnamefont
  {Reyssat}}\ and\ \bibinfo {author} {\bibfnamefont {L.}~\bibnamefont
  {Mahadevan}},\ }\href {\doibase 10.1098/rsif.2009.0184} {\bibfield  {journal}
  {\bibinfo  {journal} {Journal of the Royal Society, Interface}\ }\textbf
  {\bibinfo {volume} {6}},\ \bibinfo {pages} {951} (\bibinfo {year}
  {2009})}\BibitemShut {NoStop}%
\bibitem [{\citenamefont {Aharoni}\ \emph {et~al.}(2012)\citenamefont
  {Aharoni}, \citenamefont {Abraham}, \citenamefont {Elbaum}, \citenamefont
  {Sharon},\ and\ \citenamefont {Kupferman}}]{AAESK12}%
  \BibitemOpen
  \bibfield  {author} {\bibinfo {author} {\bibfnamefont {H.}~\bibnamefont
  {Aharoni}}, \bibinfo {author} {\bibfnamefont {Y.}~\bibnamefont {Abraham}},
  \bibinfo {author} {\bibfnamefont {R.}~\bibnamefont {Elbaum}}, \bibinfo
  {author} {\bibfnamefont {E.}~\bibnamefont {Sharon}}, \ and\ \bibinfo {author}
  {\bibfnamefont {R.}~\bibnamefont {Kupferman}},\ }\href@noop {} {\bibfield
  {journal} {\bibinfo  {journal} {Physical Review Letters}\ }\textbf {\bibinfo
  {volume} {108}},\ \bibinfo {pages} {238106} (\bibinfo {year}
  {2012})}\BibitemShut {NoStop}%
\bibitem [{\citenamefont {Sydney~Gladman}\ \emph {et~al.}(2016)\citenamefont
  {Sydney~Gladman}, \citenamefont {Matsumoto}, \citenamefont {Nuzzo},
  \citenamefont {Mahadevan},\ and\ \citenamefont {Lewis}}]{Gla16}%
  \BibitemOpen
  \bibfield  {author} {\bibinfo {author} {\bibfnamefont {A.}~\bibnamefont
  {Sydney~Gladman}}, \bibinfo {author} {\bibfnamefont {E.~A.}\ \bibnamefont
  {Matsumoto}}, \bibinfo {author} {\bibfnamefont {R.~G.}\ \bibnamefont
  {Nuzzo}}, \bibinfo {author} {\bibfnamefont {L.}~\bibnamefont {Mahadevan}}, \
  and\ \bibinfo {author} {\bibfnamefont {J.~A.}\ \bibnamefont {Lewis}},\ }\href
  {http://dx.doi.org/10.1038/nmat4544} {\bibfield  {journal} {\bibinfo
  {journal} {Nature Materials}\ }\textbf {\bibinfo {volume} {15}},\ \bibinfo
  {pages} {413 EP } (\bibinfo {year} {2016})}\BibitemShut {NoStop}%
\bibitem [{\citenamefont {Mostajeran}\ \emph {et~al.}(2016)\citenamefont
  {Mostajeran}, \citenamefont {Warner}, \citenamefont {Ware},\ and\
  \citenamefont {White}}]{War16}%
  \BibitemOpen
  \bibfield  {author} {\bibinfo {author} {\bibfnamefont {C.}~\bibnamefont
  {Mostajeran}}, \bibinfo {author} {\bibfnamefont {M.}~\bibnamefont {Warner}},
  \bibinfo {author} {\bibfnamefont {T.~H.}\ \bibnamefont {Ware}}, \ and\
  \bibinfo {author} {\bibfnamefont {T.~J.}\ \bibnamefont {White}},\ }\href
  {\doibase 10.1098/rspa.2016.0112} {\bibfield  {journal} {\bibinfo  {journal}
  {Proceedings of the Royal Society of London A: Mathematical, Physical and
  Engineering Sciences}\ }\textbf {\bibinfo {volume} {472}} (\bibinfo {year}
  {2016}),\ 10.1098/rspa.2016.0112}\BibitemShut {NoStop}%
\bibitem [{\citenamefont {Aharoni}\ \emph {et~al.}(2018)\citenamefont
  {Aharoni}, \citenamefont {Xia}, \citenamefont {Zhang}, \citenamefont
  {Kamien},\ and\ \citenamefont {Yang}}]{Aha18}%
  \BibitemOpen
  \bibfield  {author} {\bibinfo {author} {\bibfnamefont {H.}~\bibnamefont
  {Aharoni}}, \bibinfo {author} {\bibfnamefont {Y.}~\bibnamefont {Xia}},
  \bibinfo {author} {\bibfnamefont {X.}~\bibnamefont {Zhang}}, \bibinfo
  {author} {\bibfnamefont {R.~D.}\ \bibnamefont {Kamien}}, \ and\ \bibinfo
  {author} {\bibfnamefont {S.}~\bibnamefont {Yang}},\ }\href {\doibase
  10.1073/pnas.1804702115} {\bibfield  {journal} {\bibinfo  {journal} {Proc.
  Nat. Aca. Sci}\ }\textbf {\bibinfo {volume} {115}},\ \bibinfo {pages} {7206}
  (\bibinfo {year} {2018})}\BibitemShut {NoStop}%
\bibitem [{\citenamefont {Warner}\ and\ \citenamefont
  {Mostajeran}(2018)}]{WM17}%
  \BibitemOpen
  \bibfield  {author} {\bibinfo {author} {\bibfnamefont {M.}~\bibnamefont
  {Warner}}\ and\ \bibinfo {author} {\bibfnamefont {C.}~\bibnamefont
  {Mostajeran}},\ }\href {\doibase 10.1098/rspa.2017.0566} {\bibfield
  {journal} {\bibinfo  {journal} {Proc. R. Soc. A}\ }\textbf {\bibinfo {volume}
  {474}},\ \bibinfo {pages} {20170566} (\bibinfo {year} {2018})}\BibitemShut
  {NoStop}%
\bibitem [{\citenamefont {Kowalski}\ \emph {et~al.}(2018)\citenamefont
  {Kowalski}, \citenamefont {Mostajeran}, \citenamefont {Godman}, \citenamefont
  {Warner},\ and\ \citenamefont {White}}]{War18}%
  \BibitemOpen
  \bibfield  {author} {\bibinfo {author} {\bibfnamefont {B.~A.}\ \bibnamefont
  {Kowalski}}, \bibinfo {author} {\bibfnamefont {C.}~\bibnamefont
  {Mostajeran}}, \bibinfo {author} {\bibfnamefont {N.~P.}\ \bibnamefont
  {Godman}}, \bibinfo {author} {\bibfnamefont {M.}~\bibnamefont {Warner}}, \
  and\ \bibinfo {author} {\bibfnamefont {T.~J.}\ \bibnamefont {White}},\ }\href
  {\doibase 10.1103/PhysRevE.97.012504} {\bibfield  {journal} {\bibinfo
  {journal} {Phys. Rev. E}\ }\textbf {\bibinfo {volume} {97}},\ \bibinfo
  {pages} {012504} (\bibinfo {year} {2018})}\BibitemShut {NoStop}%
\bibitem [{\citenamefont {K{\"{u}}pfer}\ and\ \citenamefont
  {Finkelmann}(1991)}]{Kupfer1991}%
  \BibitemOpen
  \bibfield  {author} {\bibinfo {author} {\bibfnamefont {J.}~\bibnamefont
  {K{\"{u}}pfer}}\ and\ \bibinfo {author} {\bibfnamefont {H.}~\bibnamefont
  {Finkelmann}},\ }\href {\doibase 10.1002/marc.1991.030121211} {\bibfield
  {journal} {\bibinfo  {journal} {Die Makromolekulare Chemie, Rapid
  Communications}\ }\textbf {\bibinfo {volume} {12}},\ \bibinfo {pages} {717}
  (\bibinfo {year} {1991})}\BibitemShut {NoStop}%
\bibitem [{\citenamefont {Finkelmann}\ \emph {et~al.}(2001)\citenamefont
  {Finkelmann}, \citenamefont {Nishikawa}, \citenamefont {Pereira},\ and\
  \citenamefont {Warner}}]{Finkelmann2001}%
  \BibitemOpen
  \bibfield  {author} {\bibinfo {author} {\bibfnamefont {H.}~\bibnamefont
  {Finkelmann}}, \bibinfo {author} {\bibfnamefont {E.}~\bibnamefont
  {Nishikawa}}, \bibinfo {author} {\bibfnamefont {G.}~\bibnamefont {Pereira}},
  \ and\ \bibinfo {author} {\bibfnamefont {M.}~\bibnamefont {Warner}},\ }\href
  {\doibase 10.1103/PhysRevLett.87.015501} {\bibfield  {journal} {\bibinfo
  {journal} {Physical Review Letters}\ }\textbf {\bibinfo {volume} {87}},\
  \bibinfo {pages} {015501} (\bibinfo {year} {2001})}\BibitemShut {NoStop}%
\bibitem [{\citenamefont {Warner}\ and\ \citenamefont
  {Terentjev}(2003)}]{WT03}%
  \BibitemOpen
  \bibfield  {author} {\bibinfo {author} {\bibfnamefont {M.}~\bibnamefont
  {Warner}}\ and\ \bibinfo {author} {\bibfnamefont {E.~M.}\ \bibnamefont
  {Terentjev}},\ }\href@noop {} {\emph {\bibinfo {title} {Liquid {{Crystal
  Elastomers}}}}},\ International Series of Monographs on Physics\ (\bibinfo
  {publisher} {{Oxford University Press}},\ \bibinfo {address} {Oxford},\
  \bibinfo {year} {2003})\BibitemShut {NoStop}%
\bibitem [{\citenamefont {Modes}\ \emph {et~al.}(2010)\citenamefont {Modes},
  \citenamefont {Bhattacharya},\ and\ \citenamefont {Warner}}]{Modes2010a}%
  \BibitemOpen
  \bibfield  {author} {\bibinfo {author} {\bibfnamefont {C.~D.}\ \bibnamefont
  {Modes}}, \bibinfo {author} {\bibfnamefont {K.}~\bibnamefont {Bhattacharya}},
  \ and\ \bibinfo {author} {\bibfnamefont {M.}~\bibnamefont {Warner}},\ }\href
  {\doibase 10.1098/rspa.2010.0352} {\bibfield  {journal} {\bibinfo  {journal}
  {Proceedings of the Royal Society A: Mathematical, Physical and Engineering
  Sciences}\ }\textbf {\bibinfo {volume} {467}},\ \bibinfo {pages} {1121}
  (\bibinfo {year} {2010})}\BibitemShut {NoStop}%
\bibitem [{\citenamefont {Modes}\ and\ \citenamefont
  {Warner}(2011)}]{Modes2011}%
  \BibitemOpen
  \bibfield  {author} {\bibinfo {author} {\bibfnamefont {C.~D.}\ \bibnamefont
  {Modes}}\ and\ \bibinfo {author} {\bibfnamefont {M.}~\bibnamefont {Warner}},\
  }\href {\doibase 10.1103/PhysRevE.84.021711} {\bibfield  {journal} {\bibinfo
  {journal} {Physical Review E}\ }\textbf {\bibinfo {volume} {84}},\ \bibinfo
  {pages} {021711} (\bibinfo {year} {2011})}\BibitemShut {NoStop}%
\bibitem [{\citenamefont {Aharoni}\ \emph {et~al.}(2014)\citenamefont
  {Aharoni}, \citenamefont {Sharon},\ and\ \citenamefont {Kupferman}}]{ASK14}%
  \BibitemOpen
  \bibfield  {author} {\bibinfo {author} {\bibfnamefont {H.}~\bibnamefont
  {Aharoni}}, \bibinfo {author} {\bibfnamefont {E.}~\bibnamefont {Sharon}}, \
  and\ \bibinfo {author} {\bibfnamefont {R.}~\bibnamefont {Kupferman}},\ }\href
  {\doibase 10.1103/PhysRevLett.113.257801} {\bibfield  {journal} {\bibinfo
  {journal} {Phys. Rev. Lett.}\ }\textbf {\bibinfo {volume} {113}},\ \bibinfo
  {pages} {257801} (\bibinfo {year} {2014})}\BibitemShut {NoStop}%
\bibitem [{\citenamefont {Mostajeran}(2015)}]{Mostajeran2015}%
  \BibitemOpen
  \bibfield  {author} {\bibinfo {author} {\bibfnamefont {C.}~\bibnamefont
  {Mostajeran}},\ }\href {\doibase 10.1103/PhysRevE.91.062405} {\bibfield
  {journal} {\bibinfo  {journal} {Physical Review E - Statistical, Nonlinear,
  and Soft Matter Physics}\ }\textbf {\bibinfo {volume} {91}} (\bibinfo {year}
  {2015}),\ 10.1103/PhysRevE.91.062405},\ \Eprint
  {http://arxiv.org/abs/arXiv:1505.03140v1} {arXiv:arXiv:1505.03140v1}
  \BibitemShut {NoStop}%
\bibitem [{\citenamefont {Plucinsky}\ \emph {et~al.}(2016)\citenamefont
  {Plucinsky}, \citenamefont {Lemm},\ and\ \citenamefont
  {Bhattacharya}}]{Plucinsky2016}%
  \BibitemOpen
  \bibfield  {author} {\bibinfo {author} {\bibfnamefont {P.}~\bibnamefont
  {Plucinsky}}, \bibinfo {author} {\bibfnamefont {M.}~\bibnamefont {Lemm}}, \
  and\ \bibinfo {author} {\bibfnamefont {K.}~\bibnamefont {Bhattacharya}},\
  }\href {\doibase 10.1103/PhysRevE.94.010701} {\bibfield  {journal} {\bibinfo
  {journal} {Physical Review E}\ }\textbf {\bibinfo {volume} {94}},\ \bibinfo
  {pages} {010701} (\bibinfo {year} {2016})}\BibitemShut {NoStop}%
\bibitem [{\citenamefont {Si{\'e}fert}\ \emph {et~al.}(2019)\citenamefont
  {Si{\'e}fert}, \citenamefont {Reyssat}, \citenamefont {Bico},\ and\
  \citenamefont {Roman}}]{BRRS19}%
  \BibitemOpen
  \bibfield  {author} {\bibinfo {author} {\bibfnamefont {E.}~\bibnamefont
  {Si{\'e}fert}}, \bibinfo {author} {\bibfnamefont {E.}~\bibnamefont
  {Reyssat}}, \bibinfo {author} {\bibfnamefont {J.}~\bibnamefont {Bico}}, \
  and\ \bibinfo {author} {\bibfnamefont {B.}~\bibnamefont {Roman}},\ }\href
  {\doibase 10.1038/s41563-018-0219-x} {\bibfield  {journal} {\bibinfo
  {journal} {Nature Materials}\ }\textbf {\bibinfo {volume} {18}},\ \bibinfo
  {pages} {24} (\bibinfo {year} {2019})}\BibitemShut {NoStop}%
\bibitem [{\citenamefont {Niv}\ and\ \citenamefont {Efrati}(2018)}]{NE17}%
  \BibitemOpen
  \bibfield  {author} {\bibinfo {author} {\bibfnamefont {I.}~\bibnamefont
  {Niv}}\ and\ \bibinfo {author} {\bibfnamefont {E.}~\bibnamefont {Efrati}},\
  }\href {\doibase 10.1039/C7SM01672G} {\bibfield  {journal} {\bibinfo
  {journal} {Soft Matter}\ }\textbf {\bibinfo {volume} {14}},\ \bibinfo {pages}
  {424} (\bibinfo {year} {2018})}\BibitemShut {NoStop}%
\bibitem [{SM()}]{SM}%
  \BibitemOpen
  \href@noop {} {}\bibinfo {howpublished} {See Supplemental Material at [URL]
  for additional information on the integration scheme and bounds on
  integration distances.}\BibitemShut {Stop}%
\bibitem [{\citenamefont {Rozhdestvenskii}(1962)}]{Roz62}%
  \BibitemOpen
  \bibfield  {author} {\bibinfo {author} {\bibfnamefont {B.~L.}\ \bibnamefont
  {Rozhdestvenskii}},\ }\href@noop {} {\bibfield  {journal} {\bibinfo
  {journal} {Doklady Akademii Nauk (SSSR)}\ }\textbf {\bibinfo {volume}
  {143}},\ \bibinfo {pages} {50} (\bibinfo {year} {1962})}\BibitemShut
  {NoStop}%
\bibitem [{\citenamefont {Poznyak}\ and\ \citenamefont {Shikin}(1995)}]{PS95}%
  \BibitemOpen
  \bibfield  {author} {\bibinfo {author} {\bibfnamefont {E.~G.}\ \bibnamefont
  {Poznyak}}\ and\ \bibinfo {author} {\bibfnamefont {E.~V.}\ \bibnamefont
  {Shikin}},\ }\href@noop {} {\bibfield  {journal} {\bibinfo  {journal}
  {Journal of Mathematical Sciences}\ }\textbf {\bibinfo {volume} {74}},\
  \bibinfo {pages} {1078} (\bibinfo {year} {1995})}\BibitemShut {NoStop}%
\bibitem [{\citenamefont {Hadamard}(1923)}]{Hadamard}%
  \BibitemOpen
  \bibfield  {author} {\bibinfo {author} {\bibfnamefont {J.}~\bibnamefont
  {Hadamard}},\ }\href@noop {} {\emph {\bibinfo {title} {Lectures on Cauchy's
  problem in linear partial differential equations}}}\ (\bibinfo  {publisher}
  {Yale University press},\ \bibinfo {year} {1923})\BibitemShut {NoStop}%
\bibitem [{\citenamefont {Goursat}(1923)}]{Goursat}%
  \BibitemOpen
  \bibfield  {author} {\bibinfo {author} {\bibfnamefont {E.}~\bibnamefont
  {Goursat}},\ }\href@noop {} {\emph {\bibinfo {title} {Cours d'analyse
  math{\'e}matique}}},\ Vol.~\bibinfo {volume} {3}\ (\bibinfo  {publisher}
  {Gauthier-Villars},\ \bibinfo {year} {1923})\BibitemShut {NoStop}%
\bibitem [{\citenamefont {Gemmer}\ \emph {et~al.}(2016)\citenamefont {Gemmer},
  \citenamefont {Sharon}, \citenamefont {Shearman},\ and\ \citenamefont
  {Venkataramani}}]{Gemmer2016}%
  \BibitemOpen
  \bibfield  {author} {\bibinfo {author} {\bibfnamefont {J.}~\bibnamefont
  {Gemmer}}, \bibinfo {author} {\bibfnamefont {E.}~\bibnamefont {Sharon}},
  \bibinfo {author} {\bibfnamefont {T.}~\bibnamefont {Shearman}}, \ and\
  \bibinfo {author} {\bibfnamefont {S.~C.}\ \bibnamefont {Venkataramani}},\
  }\href {\doibase 10.1209/0295-5075/114/24003} {\bibfield  {journal} {\bibinfo
   {journal} {Europhysics Letters}\ }\textbf {\bibinfo {volume} {114}},\
  \bibinfo {pages} {24003} (\bibinfo {year} {2016})},\ \Eprint
  {http://arxiv.org/abs/1601.06863} {arXiv:1601.06863} \BibitemShut {NoStop}%
\end{thebibliography}%

\newpage
\onecolumngrid


\section*{Supplementary material}

# Supplemental Material:
# Curved geometries from planar director fields - Solving the two-dimensional inverse problem

Itay Griniasty,[1] Hillel Aharoni,[1, 2] and Efi Efrati[1, *]

[1]*Department of Physics of Complex Systems, Weizmann Institute of Science, Rehovot 76100, Israel*
[2]*Department of Physics and Astronomy, University of Pennsylvania,*
*209 South 33rd Street, Philadelphia, Pennsylvania 19104, USA*
(Dated: May 23, 2019)

## THE CURVED SYSTEM OF EQUATIONS

The desired curved surface $\mathcal{M}$ is given by a 2D parametrization

$$\mathbf{R}_\mathcal{M} : \Omega \mapsto \mathcal{M}, \quad \Omega \subset \mathbb{R}^2. \tag{1}$$

The inverse problem thus translates to finding mappings between 2D domains, $r_\Omega : \mathcal{D} \mapsto \Omega$, and $\mathbf{r} : \mathcal{D} \mapsto \mathcal{S}$, such that the actuated surface is given by $\mathbf{r}_A(u,v) \equiv \mathbf{R}_\mathcal{M}(r_\Omega(u,v))$ (see Fig. 1).

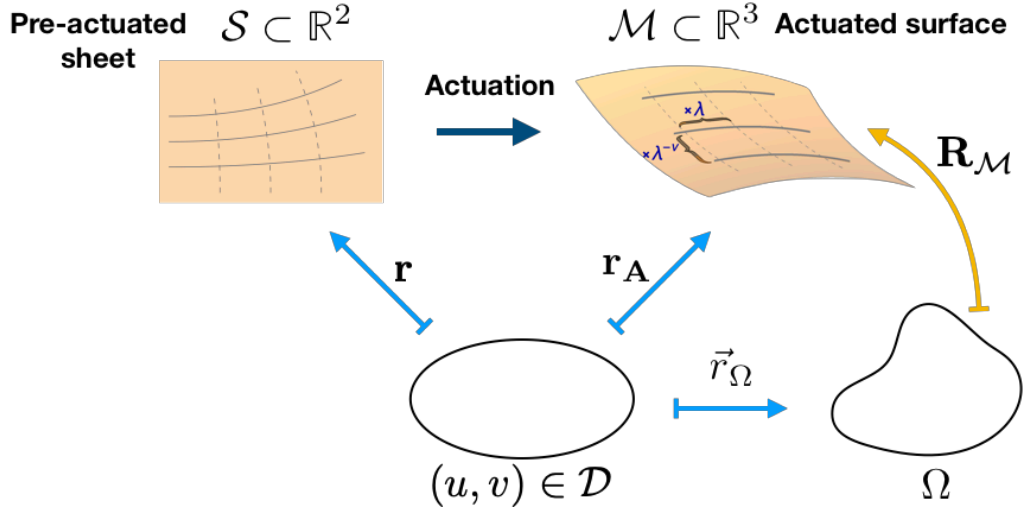

FIG. 1. The inverse problem is solved by co-propagating the bend, splay, directors and arc-lengths on the pre-actuated sheet and the curved surface along the director induced coordinates $u$ and $v$. That is a solution is found as maps $\mathbf{r}$ and $\mathbf{r}_A$ from $\mathcal{D}$ to $\mathcal{S}$ and $\mathcal{M}$. The curved actuated surface is parametrized through $\mathbf{R}_\mathcal{M}$ on a 2D domain $\Omega$. The map $r_\Omega$ is used to integrate the intrinsic equations on the curved actuated surface such that $\mathbf{r}_A = \mathbf{R}_\mathcal{M} \circ r_\Omega$.

The System Of Equations (SOE), described in the body of the letter, requires knowledge of the Gaussian curvature of the desired surface, $K_\mathcal{M}$. With the exception of a constant curvature, this implies that integrating the SOE requires keeping track of the position in the activated sheet, $\mathbf{r}_A$. This in turn is not inherited from $\mathbf{r}$ and must be integrated separately. We therefore have a Flat System Of Equations (FSOE), summarized in the left column of table I, for integrating $\hat{\mathbf{n}}, \mathbf{r}, \alpha, \beta, b$ and $s$, and a corresponding Curved System Of Equations (CSOE), summarized in the right column of table I, for the propagation of $\hat{\mathbf{n}}_A$ and $\mathbf{r}_A$. The curved arc-length parameters and geodesic curvatures, $\alpha_A, \beta_A, b_A$ and $s_A$, are related to their flat counterparts algebraically.

The embedding $\mathbf{R}_\mathcal{M}$ endows $\Omega$ with the induced metric

$$dl_A^2 = \sum_{ij=1}^{2} a_{ij} dr_\Omega^i dr_\Omega^j, \quad a_{ij} = \frac{\partial \mathbf{R}_\mathcal{M}}{\partial r_\Omega^i} \cdot \frac{\partial \mathbf{R}_\mathcal{M}}{\partial r_\Omega^j}. \tag{2}$$

The actuated director field $\hat{\mathbf{n}}_A$ is naturally endowed with an intrinsic form $\hat{n}_\Omega$, such that $\hat{\mathbf{n}}_A = (\hat{n}_\Omega \cdot \nabla_\Omega) \mathbf{R}_\mathcal{M}$, and the propagation of $r_\Omega^i$ along $u$ and $v$ is given by

$$\partial_u r_\Omega^i = \alpha_A \hat{n}_\Omega^i, \quad \partial_v r_\Omega^i = \beta_A \hat{n}_{\Omega\perp}^i. \tag{3}$$

|   | Flat | Curved |
|---|------|--------|
| u | $\frac{1}{\alpha}\partial_u \mathbf{r} = \hat{\mathbf{n}}$     (I) | $\frac{1}{\alpha_A}\partial_u r_\Omega^i = \hat{n}_\Omega^i$     (II) |
|   | $\frac{1}{\alpha}\partial_u \hat{\mathbf{n}} = b\,\hat{\mathbf{n}}_\perp$ | $\frac{1}{\alpha_A}\partial_u \hat{n}_\Omega^i = b_A\,\hat{n}_{\Omega\perp}^i - \Gamma^i_{jk}\hat{n}_\Omega^j \hat{n}_\Omega^k$ |
|   | $\frac{1}{\alpha}\partial_u(\log\beta) = s$ | $\beta_A = \lambda^{-\nu}\beta$ |
|   | $\frac{1}{\alpha}\partial_u s = -s^2 - \frac{K_{\mathcal{M}}}{\lambda^{-2}-\lambda^{2\nu}}$ | $s_A = \lambda^{-1} s$ |
| v | $\frac{1}{\beta}\partial_v \mathbf{r} = \hat{\mathbf{n}}_\perp$     (III) | $\frac{1}{\beta_A}\partial_v r_\Omega^i = \hat{n}_{\Omega\perp}^i$     (IV) |
|   | $\frac{1}{\beta}\partial_v \hat{\mathbf{n}} = s\,\hat{\mathbf{n}}_\perp$ | $\frac{1}{\beta_A}\partial_v \hat{n}_\Omega^i = s_A\,\hat{n}_{\Omega\perp}^i - \Gamma^i_{jk}\hat{n}_{\Omega\perp}^j \hat{n}_\Omega^k$ |
|   | $\frac{1}{\beta}\partial_v(\log\alpha) = -b$ | $\alpha_A = \lambda\alpha$ |
|   | $\frac{1}{\beta}\partial_v b = b^2 - \frac{K_{\mathcal{M}}}{\lambda^{-2}-\lambda^{2\nu}}$ | $b_A = \lambda^\nu b$ |

TABLE I. The SOE describing a director field $\hat{\mathbf{n}}$ on a flat sheet and $\hat{n}_\Omega$ on a curved surface, such that upon actuation the flat sheet deforms into the pre-specified curved surface. Cell I: Propagation on the flat sheet along $\hat{\mathbf{n}}$. Cell II: Propagation on the curved surface along $\hat{n}_\Omega$. Cell III: Propagation on the flat sheet along $\hat{\mathbf{n}}_\perp$. Cell IV: Propagation on the curved surface along $\hat{n}_{\Omega\perp}$.

The change of coordinates to $(u,v)$ allows writing the metric (2) on $\Omega$ in the form $dl_A^2 = \alpha_A^2 du^2 + \beta_A^2 dv^2$. The propagation of the actuated director field $\hat{n}_\Omega$ is given by the actuated bend and splay

$$b_A = \hat{n}_{\Omega\perp} \cdot (\hat{n}_\Omega \cdot \nabla)\hat{n}_\Omega, \quad s_A = \hat{n}_{\Omega\perp} \cdot (\hat{n}_{\Omega\perp} \cdot \nabla)\hat{n}_\Omega. \tag{4}$$

As the metric $a_{ij}$ is generally non-Euclidean, the derivatives of $\hat{n}_\Omega$ along and across itself, parametrized by $u$ and $v$, include a contribution from the non-flat connection $\Gamma^i_{jk}$:

$$\frac{1}{\alpha_A}\partial_u \hat{n}_\Omega^i = b_A\,\hat{n}_{\Omega\perp}^i - \Gamma^i_{jk}\hat{n}_\Omega^j \hat{n}_\Omega^k, \quad \frac{1}{\beta_A}\partial_v \hat{n}_\Omega^i = s_A\,\hat{n}_{\Omega\perp}^i - \Gamma^i_{jk}\hat{n}_{\Omega\perp}^j \hat{n}_\Omega^k. \tag{5}$$

The bend and splay can be given also in terms of the arc length parameters [1]:

$$b_A = -\frac{\partial_v \alpha_A}{\alpha_A \beta_A}, \quad s_A = \frac{\partial_u \beta_A}{\alpha_A \beta_A}. \tag{6}$$

For completeness we repeat the derivation of these equations. Exchanging the directional derivatives along and across the director by derivatives with respect to $u$ and $v$, and noting that $\hat{n}_\Omega^i = \frac{1}{\alpha_A}\partial_u r_\Omega^i$ and $\hat{n}_{\Omega\perp}^i = \frac{1}{\beta_A}\partial_v r_\Omega^i$, we rewrite equation (4) as

$$b_A = \frac{1}{\beta_A}\partial_v r_\Omega \cdot \frac{1}{\alpha_A}\partial_u \frac{1}{\alpha_A}\partial_u r_\Omega = \frac{1}{\alpha_A^2 \beta_A}\Gamma_{vuu} = -\frac{\partial_v \alpha_A}{\alpha_A \beta_A},$$

$$s_A = \frac{1}{\beta_A}\partial_v r_\Omega \cdot \frac{1}{\beta_A}\partial_v \frac{1}{\alpha_A}\partial_u r_\Omega = \frac{1}{\beta_A^2 \alpha_A}\Gamma_{vuv} = \frac{\partial_u \beta_A}{\alpha_A \beta_A},$$

where $\Gamma$ is the Levi-Civita connection associated with the metric $dl_A^2 = \alpha_A^2 du^2 + \beta_A^2 dv^2$.

The same derivation holds also for the flat bend and splay. Recalling that the actuated arc-length parameters are related to their flat value by the anisotropic deformation factors

$$\alpha_A = \lambda\alpha, \quad \beta_A = \lambda^{-\nu}\beta, \tag{7}$$

we find that the actuated bend and splay are also algebraically related to their flat counterparts

$$b_A = \lambda^\nu b, \quad s_A = \lambda^{-1} s. \tag{8}$$

Equations (3), (5), (7) and (8), composing the CSOE, determine $\mathbf{r}_A(u,v)$ and $K(u,v)$ given the simultaneous solution of the FSOE. The CSOE is trivially integrable as $u,v$ are simply a change of coordinates from $r_\Omega^i$. We have thus found a complete and integrable set of equations, collected in table I, describing flat director fields that deform on actuation into a predetermined curved surface.

## CURVED PRE-ACTUATED SHEET

The framework presented above allows us to solve a generalized form of the inverse problem: "What is the director orientation field $\hat{\mathbf{n}}(\mathbf{r})$, *on an initially curved sheet* $\mathcal{S}$, that will result in a desired surface geometry, $\mathcal{M}$, upon actuation?".

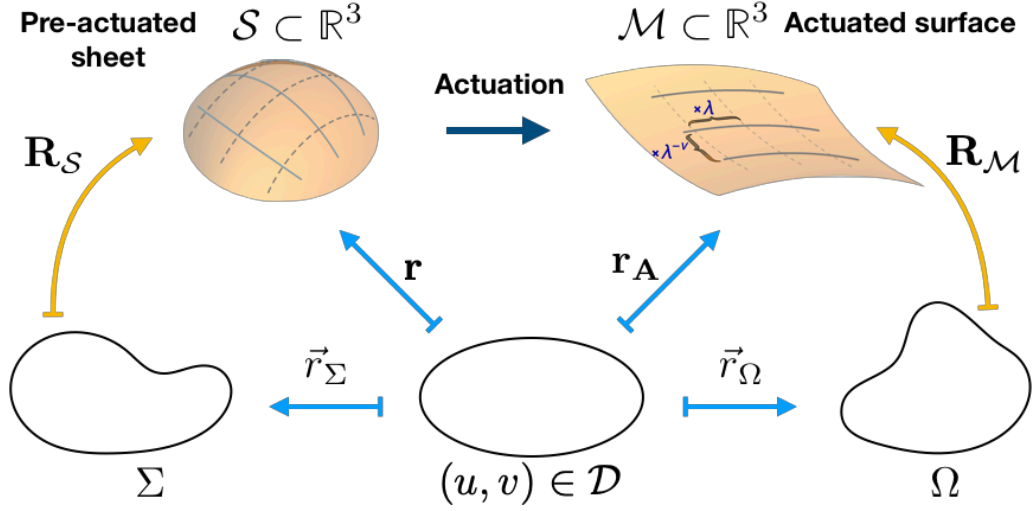

FIG. 2. The generalized inverse problem between a curved pre-actuated sheet and an actuated surface is solved by co-propagating the bend, splay, directors and arc-lengths on the pre-actuated sheet and the curved surface along the director induced coordinates $u$ and $v$. That is a solution is found as maps $\mathbf{r}$ and $\mathbf{r}_A$ from $\mathcal{D}$ to $\mathcal{S}$ and $\mathcal{M}$. The curved pre-actuated sheet and actuated surface are parametrized through $\mathbf{R}_\mathcal{S}$ and $\mathbf{R}_\mathcal{M}$ on 2D domains $\Sigma$ and $\Omega$ respectively. The maps $r_\Sigma$ and $r_\Omega$ are used to integrate the intrinsic system of equations such that $\mathbf{r} = \mathbf{R}_\mathcal{S} \circ r_\Sigma$ and $\mathbf{r}_A = \mathbf{R}_\mathcal{M} \circ r_\Omega$.

To answer this problem consider an initial curved sheet $\mathcal{S}$ given by a two dimensional parametrization

$$\mathbf{R}_\mathcal{S} : \Sigma \mapsto \mathcal{S}, \quad \Sigma \subset \mathbb{R}^2, \quad \mathcal{S} \subset \mathbb{R}^3. \tag{9}$$

For the flat sheet considered in the body of the letter the embedding is the trivial one $\mathbf{R}_\mathcal{S} : (x, y) \mapsto (x, y, 0)$. Similarly to (2) the embedding $\mathbf{R}_\mathcal{S}$ endows $\Sigma$ with an induced metric $\widetilde{a}_{ij}$, curvature $K_\mathcal{S}$ and an intrinsic form of the director field $\hat{n}_\Sigma$. The propagation along $u$ and $v$ is calculated through the intrinsic mapping $r_\Sigma : \mathcal{D} \mapsto \Sigma$, such that $\mathbf{r}(u,v) \equiv \mathbf{R}_\mathcal{S}(r_\Sigma(u,v))$ (see Fig. 2).

The compatibility equation on $\mathcal{S}$ reads

$$-K_\mathcal{S} = \left(s^2 + \frac{1}{\alpha}\partial_u s\right) + \left(b^2 - \frac{1}{\beta}\partial_v b\right), \tag{10}$$

from which we derive generalized propagation equations for the bend and splay

$$\frac{1}{\beta}\partial_v b = b^2 - \frac{K_\mathcal{M} - \lambda^{-2}K_\mathcal{S}}{\lambda^{-2} - \lambda^{2\nu}}, \quad \frac{1}{\alpha}\partial_u s = -s^2 + \frac{-K_\mathcal{M} + K_\mathcal{S}\lambda^{2\nu}}{\lambda^{-2} - \lambda^{2\nu}}. \tag{11}$$

The propagation of the director on the curved sheet $\mathcal{S}$, similarly to Eq. (5), includes a correction due to the connection $\widetilde{\Gamma}^i_{jk}$,

$$\frac{1}{\alpha}\partial_u \hat{n}^i_\Sigma = b\,\hat{n}^i_{\Sigma\perp} - \widetilde{\Gamma}^i_{jk}\hat{n}^j_\Sigma \hat{n}^k_\Sigma, \quad \frac{1}{\beta}\partial_v \hat{n}^i_\Sigma = s\,\hat{n}^i_{\Sigma\perp} - \widetilde{\Gamma}^i_{jk}\hat{n}^j_{\Sigma\perp}\hat{n}^k_\Sigma. \tag{12}$$

We thus arrive at an equivalent system of equations describing the generalized inverse problem, summarized in Table II.

|  | Pre-Actuated | Actuated |
|---|---|---|
| u | $\frac{1}{\alpha}\partial_u r^i_\Sigma = \hat{n}^i_\Sigma$  (I) | $\frac{1}{\alpha_A}\partial_u r^i_\Omega = \hat{n}^i_\Omega$  (II) |
|  | $\frac{1}{\alpha}\partial_u \hat{n}^i_\Sigma = b\,\hat{n}^i_{\Sigma\perp} - \widetilde{\Gamma}^i_{jk}\hat{n}^j_\Sigma \hat{n}^k_\Sigma$ | $\frac{1}{\alpha_A}\partial_u \hat{n}^i_\Omega = b_A\,\hat{n}^i_{\Omega\perp} - \Gamma^i_{jk}\hat{n}^j_\Omega \hat{n}^k_\Omega$ |
|  | $\frac{1}{\alpha}\partial_u(\log\beta) = s$ | $\beta_A = \lambda^{-\nu}\beta$ |
|  | $\frac{1}{\alpha}\partial_u s = -s^2 - \frac{K_\mathcal{M}-K_\mathcal{S}/\lambda^2}{\lambda^{-2}-\lambda^{2\nu}}$ | $s_A = \lambda^{-1} s$ |
| v | $\frac{1}{\beta}\partial_v r^i_\Sigma = \hat{n}^i_{\Sigma\perp}$  (III) | $\frac{1}{\beta_A}\partial_v r^i_\Omega = \hat{n}^i_{\Omega\perp}$  (IV) |
|  | $\frac{1}{\beta}\partial_v \hat{n}^i_\Sigma = s\,\hat{n}^i_{\Sigma\perp} - \widetilde{\Gamma}^i_{jk}\hat{n}^j_{\Sigma\perp}\hat{n}^k_\Sigma$ | $\frac{1}{\beta_A}\partial_v \hat{n}^i_\Omega = s_A\,\hat{n}^i_{\Omega\perp} - \Gamma^i_{jk}\hat{n}^j_{\Omega\perp}\hat{n}^k_\Omega$ |
|  | $\frac{1}{\beta}\partial_v(\log\alpha) = -b$ | $\alpha_A = \lambda\alpha$ |
|  | $\frac{1}{\beta}\partial_v b = b^2 - \frac{K_\mathcal{M}-K_\mathcal{S}\lambda^{2\nu}}{\lambda^{-2}-\lambda^{2\nu}}$ | $b_A = \lambda^\nu b$ |

TABLE II. The system of equations describing a director field $\hat{n}_\Sigma$ on an initially curved sheet and $\hat{n}_\Omega$ on a curved desired surface, such that the sheet deforms upon actuation into the pre-specified curved surface. Cell I: Propagation on the pre-actuated sheet along $\hat{n}_\Sigma$. Cell II: Propagation on the actuated surface along $\hat{n}_\Omega$. Cell III: Propagation on the pre-actuated sheet along $\hat{n}_{\Sigma\perp}$. Cell IV: Propagation on the actuated surface along $\hat{n}_{\Omega\perp}$.

## BOUND ON DISTANCE

Consider a surface with positive curvature $K_\mathcal{M} \geq K_0$. Denote $\omega = \sqrt{\frac{K_0}{\lambda^{-2}-\lambda^{2\nu}}}$, and the derivative along the director, $\hat{\mathbf{n}}\cdot\nabla$, by $\frac{d}{d\sigma}$ where $\sigma$ is an arc-length parameter along $\hat{\mathbf{n}}$ integral lines ($u$-parametric curves). The splay's propagation equation satisfies

$$\left.\frac{ds}{d\sigma}\right|_v = -s^2 - \frac{K_\mathcal{M}}{\lambda^{-2}-\lambda^{2\nu}} \leq -s^2 - \omega^2. \tag{13}$$

Define

$$s_0 = \omega\tan(\omega(\xi_0-\sigma)), \qquad \xi_0 = \frac{1}{\omega}\arctan(s(0)/\omega) \leq \frac{\pi}{2\omega}, \tag{14}$$

such that $s_0(0) = s(0)$, and $\left.\frac{ds_0}{d\sigma}\right|_v = -s_0^2 - \omega^2$. The above equations imply that for all $\sigma$ with $s_0(\sigma) = s(\sigma)$, $\left.\frac{ds_0}{d\sigma}\right|_v(\sigma) \geq \frac{ds}{d\sigma}(\sigma)$, and as $s_0(0) = s(0)$, $s_0 \geq s$. Recall that $\frac{d(\log\beta)}{d\sigma} = s$, hence

$$\beta = \beta_0 e^{\int s} \leq \beta_0 e^{\int s_0} \tag{15}$$

$$\Rightarrow \beta \leq \beta_0 \cos(\omega(\xi_0-\sigma)) \tag{16}$$

Thus $\beta(\sigma^*) = 0$ for $\sigma^* \leq \xi_0 + \frac{\pi}{2\omega} \leq \frac{\pi}{\omega}$.

Let $\sigma$ denote now the arc length parameter along $\hat{\mathbf{n}}_\perp$ integral lines ($v$ - parametric curves). The derivative along $\hat{\mathbf{n}}_\perp$ on such lines is again bound by:

$$\left.\frac{db}{d\sigma}\right|_v \leq b^2 - \omega^2, \tag{17}$$

following a similar argument:

$$b \leq b_0 \equiv \omega\tanh(\omega(\eta_0-\sigma)) \tag{18}$$

$$\eta_0 = \frac{1}{\omega}\text{arctanh}(b(0)/\omega) \tag{19}$$

$$\left.\frac{d(\log\alpha)}{d\sigma}\right|_v = -b \Rightarrow \alpha = \alpha_0 e^{-\int b} \geq \alpha_0 e^{-\int b_0} \tag{20}$$

$$\Rightarrow \alpha \geq \alpha_0 \cosh(\omega(\eta_0-\sigma)) > 0. \tag{21}$$

Applying this result to solving for a director that would actuate into a sphere of radius $R$ we find that the actuated director curve extends no longer than $\lambda\pi\omega < \pi R$, i.e. a director curve always spans less than half the circumference of a sphere, and an actuated smooth director field cannot correspond to a single cover of an entire sphere.

## GRAIN BOUNDARY IN A FACE'S SYMMETRY AXIS

The director field found numerically in figure 1 of [2] displays what appears to be a grain boundary along the right-left symmetry axis of the face. For any symmetric director field the symmetry axis is either a $\hat{\mathbf{n}}_A$-line or a $\hat{\mathbf{n}}_{A\perp}$-line. Consider a director field where the symmetry axis is a director-line, like the one found in [2]. Such a symmetric smooth director field is thus constrained along the symmetry axis by the bounding equation on $\beta$:

$$\left.\frac{d^2\beta}{d\sigma^2}\right|_v = -\frac{K_{\mathcal{M}}}{\lambda^{-2}-\lambda^{2\nu}}\beta. \quad (22)$$

The face's Gaussian curvature peaks at its nose. Integrating $\beta$ along the symmetry axis from the nose, with an initial condition $d\beta/d\sigma|_v = 0$, that is $s = 0$, the director field arrives at a singularity, $\beta = 0$, both in the direction of the forehead and the chin. For any other initial condition the director field arrives at the singularity faster either in the forehead's direction, or the chin's. We thus find that a symmetric smooth director field aligned with the symmetry axis cannot cover the entire face, but a symmetric director field with a grain boundary might.

---

* efi.efrati@weizmann.ac.il
[1] I. Niv and E. Efrati, Soft Matter **14**, 424 (2018).
[2] H. Aharoni, Y. Xia, X. Zhang, R. D. Kamien, and S. Yang, Proc. Nat. Aca. Sci **115**, 7206 (2018).



\end{document}